\documentstyle[twoside,epsf]{article}

\catcode`\@=11
\long\def\@makefntext#1{
\protect\noindent \hbox to 3.2pt {\hskip-.9pt
$^{{\eightrm\@thefnmark}}$\hfil}#1\hfill}		

\def\@makefnmark{\hbox to 0pt{$^{\@thefnmark}$\hss}}	
	
\def\ps@myheadings{\let\@mkboth\@gobbletwo
\def\@oddhead{\hbox{}
\rightmark\hfil\eightrm\thepage}
\def\@oddfoot{}\def\@evenhead{\eightrm\thepage\hfil
\leftmark\hbox{}}\def\@evenfoot{}
\def\sectionmark##1{}\def\subsectionmark##1{}}

\oddsidemargin=\evensidemargin
\newcounter{sectionc}\newcounter{subsectionc}\newcounter{subsubsectionc}
\renewcommand{\section}[1] {\vspace{12pt}\addtocounter{sectionc}{1}
\setcounter{subsectionc}{0}\setcounter{subsubsectionc}{0}\noindent
	{\tenbf\thesectionc. #1}\par\vspace{5pt}}
\renewcommand{\subsection}[1] {\vspace{12pt}\addtocounter{subsectionc}{1}
	\setcounter{subsubsectionc}{0}\noindent
	{\bf\thesectionc.\thesubsectionc. {\kern1pt \bfit #1}}\par\vspace{5pt}}
\renewcommand{\subsubsection}[1] {\vspace{12pt}\addtocounter{subsubsectionc}{1}
	\noindent{\tenrm\thesectionc.\thesubsectionc.\thesubsubsectionc.
	{\kern1pt \tenit #1}}\par\vspace{5pt}}
\newcommand{\nonumsection}[1] {\vspace{12pt}\noindent{\tenbf #1}
	\par\vspace{5pt}}

\newcommand{\textlineskip}{\baselineskip=13pt}

\def\eightcirc{
\begin{picture}(0,0)
\put(4.4,1.8){\circle{6.5}}
\end{picture}}
\def\eightcopyright{\eightcirc\kern2.7pt\hbox{\eightrm c}}

\def\abstracts#1#2#3{{
	\centering{\begin{minipage}{4.5in}\baselineskip=10pt\footnotesize
	\parindent=0pt #1\par
	\parindent=15pt #2\par
	\parindent=15pt #3
	\end{minipage}}\par}}

\def\keywords#1{{
	\centering{\begin{minipage}{4.5in}\baselineskip=10pt\footnotesize
	{\footnotesize\it Keywords}\/: #1
	 \end{minipage}}\par}}

\renewenvironment{thebibliography}[1]
	{\frenchspacing
	 \ninerm\baselineskip=11pt
	 \begin{list}{\arabic{enumi}.}
        {\usecounter{enumi}\setlength{\parsep}{0pt}
	 \setlength{\leftmargin 12.7pt}{\rightmargin 0pt} 
         \setlength{\itemsep}{0pt} \settowidth
	{\labelwidth}{#1.}\sloppy}}{\end{list}}

\newcounter{itemlistc}
\newcounter{romanlistc}
\newcounter{alphlistc}
\newcounter{arabiclistc}

\def\@citex[#1]#2{\if@filesw\immediate\write\@auxout
	{\string\citation{#2}}\fi
\def\@citea{}\@cite{\@for\@citeb:=#2\do
	{\@citea\def\@citea{,}\@ifundefined
	{b@\@citeb}{{\bf ?}\@warning
	{Citation `\@citeb' on page \thepage \space undefined}}
	{\csname b@\@citeb\endcsname}}}{#1}}

\newif\if@cghi
\def\cite{\@cghitrue\@ifnextchar [{\@tempswatrue
	\@citex}{\@tempswafalse\@citex[]}}
\def\citelow{\@cghifalse\@ifnextchar [{\@tempswatrue
	\@citex}{\@tempswafalse\@citex[]}}
\def\@cite#1#2{{$\null^{#1}$\if@tempswa\typeout
	{IJCGA warning: optional citation argument
	ignored: `#2'} \fi}}

\def\@refcitex[#1]#2{\if@filesw\immediate\write\@auxout
	{\string\citation{#2}}\fi
\def\@citea{}\@refcite{\@for\@citeb:=#2\do
	{\@citea\def\@citea{, }\@ifundefined
	{b@\@citeb}{{\bf ?}\@warning
	{Citation `\@citeb' on page \thepage \space undefined}}
	\hbox{\csname b@\@citeb\endcsname}}}{#1}}

\def\@refcite#1#2{{#1\if@tempswa\typeout
        {IJCGA warning: optional citation argument
	ignored: `#2'} \fi}}

\def\refcite{\@ifnextchar[{\@tempswatrue
	\@refcitex}{\@tempswafalse\@refcitex[]}}

\def\pmb#1{\setbox0=\hbox{#1}
	\kern-.025em\copy0\kern-\wd0
	\kern.05em\copy0\kern-\wd0
	\kern-.025em\raise.0433em\box0}

\def\fnt#1#2{\footnotetext{\kern-.3em
	{$^{\mbox{\scriptsize #1}}$}{#2}}}

\def\runninghead#1#2{\pagestyle{myheadings}
\markboth{{\protect\footnotesize\it{\quad #1}}\hfill}
{\hfill{\protect\footnotesize\it{#2\quad}}}}
\font\tenrm=cmr10
\font\tenit=cmti10
\font\tenbf=cmbx10
\font\bfit=cmbxti10 at 10pt
\font\ninerm=cmr9

\font\eightrm=cmr8

\def\qed{\hbox{${\vcenter{\vbox{			
   \hrule height 0.4pt\hbox{\vrule width 0.4pt height 6pt
   \kern5pt\vrule width 0.4pt}\hrule height 0.4pt}}}$}}

\begin{document}

\newpage

\runninghead{Comment on Demystifying}
{A. Gallegos and H.C. Rosu}

\normalsize\textlineskip
\thispagestyle{empty}
\setcounter{page}{1}


\begin{center} Mod. Phys. Lett. A 33$_{24}$ (2018) 1875001 (3 pages)\\  	arXiv:1806.11139  \end{center}

\vspace*{0.88truein}

\bigskip
\centerline{\bf  Comment on Demystifying the constancy of the Ermakov-Lewis invariant}
\centerline{\bf for a time-dependent oscillator}
\vspace*{0.035truein}
\vspace*{0.37truein}
\vspace*{10pt}
\centerline{\footnotesize A. Gallegos$^1$ and H.C. Rosu$^2$\footnote{Corresponding author}}
\vspace*{0.015truein}
\centerline{\footnotesize $^1$ Departamento de Ciencias Exactas y Tecnolog\'{\i}a,\ Centro Universitario de los Lagos, Universidad de Guadalajara}
\centerline{\footnotesize Enrique D\'{\i}az de Le\'{o}n 1144, Col. Paseos de la Monta\~{n}a, Lagos de Moreno, Jalisco, Mexico}
\centerline{\footnotesize $^2$ IPICyT, Instituto Potosino de Investigacion Cientifica y Tecnologica}
 \centerline{\footnotesize Camino a la presa San Jos\'e, Col. Lomas 4a. Secci\'on, 78231 San Luis Potos\'{\i}, S.L.P., Mexico}
\vspace*{0.225truein}
\centerline{\footnotesize gallegos@culagos.ugd.mx}
\centerline{\footnotesize hcr@ipicyt.edu.mx}
\vspace*{0.21truein}
\abstracts{We show that a simple modification of the Lagrangian proposed by Padmanabhan in the paper [Mod. Phys. Lett. A {\bf 33}, 1830005 (2018)]
leads to the most general dynamical invariant in 
[Ray and Reid, Phys. Lett. A \textbf{71}, 317 (1979)].}{}{}
\vspace*{10pt}
\textlineskip                  
\vspace*{12pt}                 
\vspace*{1pt}\textlineskip	
\vspace*{-0.5pt}

\noindent

In an interesting paper, Padmanabhan\cite{Padmanabhan} shows that the Lagrangian $L_q$ of a parametric oscillator $q(t)$ with time-dependent mass can be transformed up to a total time derivative into the Lagrangian $L_Q$ of a harmonic oscillator of unit mass and constant frequency $\Omega$, where $\Omega^2$ enters as the coupling constant of the inverse cubic nonlinearity in the associated Ermakov-Milne-Pinney equation of the original oscillator $q(t)$. Using this connection, Padmanabhan concludes that the energy of the conservative oscillator $Q(t)$ is precisely the Ermakov-Lewis invariant of the parametric oscillator $q(t)$. He goes a step further and presents a generalization of the procedure valid for any (anharmonic) potential $V(Q)$, not only for the harmonic one, $V(Q)=(1/2)\Omega^2Q^2$.
In this more general case, the nonlinear coupling constant is a $Q$-running coupling constant given by $F(Q)=V'(Q)/Q$, while in the harmonic oscillator case $F(Q)$ reduces to a constant. For this more general case, Padmanabhan shows again that the energy corresponds to a case of Ray-Reid invariants.\cite{RayReid1979}

However, Padmanabhan does not present the most general case of Ray-Reid invariants, as expressed by equation (15) in their paper, which refers to the system of nonlinear oscillator equations of the form
\begin{eqnarray}
\ddot{x}+\omega^2(t)x&=(1/\rho x^2)g(\rho/x)~, \label{rr1}\\
\ddot{\rho}+\omega^2(t)\rho&=(1/\rho^2 x)h(x/\rho)~, \label{rr2}
\end{eqnarray}
where $g$ and $h$ are arbitrary functions.

Our aim here is to fill in this shortcoming.
For this, we propose the more general Lagrangian $L_Q$ given by
\begin{equation}
L_Q=\frac {1}{2} Q'^2-V(Q)-W\left(\frac{1}{Q}\right),
\label{LQ}
\end{equation}
\noindent where $'$ refers to the time derivative with respect to $\tau$ defined next. This Lagrangian has the additional potential term $W$ with respect to that of Padmanabhan. Performing the transformations,\cite{t1,t2,t3,t4}
\begin{equation}
q=fQ, \ \ \ \ \ dt=mf^2d\tau~,
\end{equation}
\noindent the Lagrangian in the new variables, up to the total derivative $-\frac{1}{2}\frac{d}{dt}(\frac{q^2}{f} m\dot{f})$ which does not change the equation of motion,\cite{GF} is
\begin{equation}
\tilde{L}_q=\frac{1}{2}m \dot{q}^2+\frac{1}{2} \frac{q^2}{f} \frac{d}{dt} (m\dot{f}) - \frac{V\left(\frac{q}{f}\right)}{mf^2} - \frac{W\left(\frac{f}{q}\right)}{mf^2}~.
\label{Lq}
\end{equation}
\noindent If we apply the Euler-Lagrange procedure, we obtain the equation of motion
\begin{equation}
\frac{d}{dt} (m\dot{q})=\frac{q}{f} \frac{d}{dt} (m\dot{f}) - \frac{V'\left(\frac{q}{f}\right)}{mf^3} + \frac{W'\left(\frac{f}{q}\right)}{mq^2f}~,
\end{equation}
where now $'$ represents the derivative with respect to its argument and the point stands for the time derivative with respect to $t$.

\medskip

If we now impose the equivalent of the first equation (\ref{rr1}) from the Ray-Reid type system 
\begin{equation}\label{equiv1}
\frac{d}{dt} (m\dot{q}) + m\tilde{\omega}^2(t)q = \frac{G\left(\frac{f}{q}\right)}{mq^3}~,
\end{equation}
and denote
\begin{equation}
F(Q)=\frac{V'(Q)}{Q}, \ \ \ \ \ G\left(\frac{1}{Q}\right)=W'\left(\frac{1}{Q}\right)Q~,
\end{equation}
we obtain the equivalent of the second equation (\ref{rr2}) from the Ray-Reid system
\begin{equation}
\frac{d}{dt} (m\dot{f}) + m\tilde{\omega}^2(t)f = \frac{F\left(\frac{q}{f}\right)}{mf^3}~.
\end{equation}
This equivalence can be seen for example by using the change of variables $x=q\sqrt{m}$ and $\rho=f\sqrt{m}$ in the latter equation, which becomes
$$
\ddot{\rho}+\omega^2(t)\rho=\frac{F\left(\frac{x}{\rho}\right)}{\rho^3},
$$
where
$$
\omega^2(t)=\frac{1}{4}m^{-2}\dot{m}^2-\frac{1}{2}m^{-1}\ddot{m}+\tilde{\omega}^2(t)~.
$$
If one chooses $F(x/\rho)=h(x/\rho)/(x/\rho)$, then one obtains (\ref{rr2}). The equivalence of (\ref{equiv1}) and (\ref{rr1}) is proved in the same way.

\noindent The conserved energy associated to the Lagrangian (\ref{LQ}) is then of the same form as the most general invariant
proposed by Ray and Reid in the transformed system
\begin{equation}
E=\frac{1}{2} Q'^2+V(Q)+W\left(\frac{1}{Q}\right)= \frac{1}{2} m^2 (\dot{q}f-q\dot{f})^2 + \int{\frac{q}{f} Fd\left( \frac{q}{f}\right)} + \int{\frac{f}{q} Gd\left( \frac{f}{q}\right)}~.
\end{equation}
This also shows that energy type invariants can be associated to special pairs of Ermakov-Milne-Pinney type equations with nonlinear running coupling constants relating (an)harmonic and singular oscillators.

\nonumsection{Acknowledgements}

\noindent
The authors thank the anonymous reviewer for constructive and helpful remarks.


\end{document}